\def\gsim{\mathrel{\rlap{\lower4pt\hbox{\hskip1pt$\sim$}}
    \raise1pt\hbox{$>$}}}         
\def\lsim{\mathrel{\rlap{\lower4pt\hbox{\hskip1pt$\sim$}}
    \raise1pt\hbox{$<$}}}         

\documentclass{PoS}

\usepackage{url}

\title{PDF4LHC recommendations for Run II}

\ShortTitle{PDF4LHC recommendations for Run II}

\author{\speaker{Juan Rojo}~\footnote{Presented on behalf of the PDF4LHC Working Group.}\\
       University of Oxford\\
       E-mail: \email{juan.rojo@physics.ox.ac.uk}}

\abstract{The interpretation of LHC measurements requires
a careful estimate of
various sources of uncertainties that affect
theoretical calculations.
In this contribution, we present the PDF4LHC Working Group recommendations
for the usage of sets of parton distribution functions (PDFs)
at the LHC Run II.
We review the construction and validation of the PDF4LHC15 combined sets, and study
some of their phenomenological implications.
We also address some recent criticism of these recommendations.
}

\FullConference{XXIV International Workshop on Deep-Inelastic Scattering and Related Subjects\\
         11-15 April, 2016\\
         DESY Hamburg, Germany}

\begin{document}

\paragraph{Why a recommendation?}
Given the ever-increasing precision of LHC measurements, careful
assessment of theoretical uncertainties in LHC cross-sections
is of utmost importance.
One of the dominant sources arises from our imperfect knowledge of
the structure of the proton, encoded by the parton distribution
functions (PDFs)~\cite{Forte:2013wc}, as well as of related physical parameters such
as the strong coupling $\alpha_s$ or the charm mass $m_c$.
Quantifying the total PDF$+\alpha_s$ uncertainties is of high importance
in a number of LHC applications: a prime example is the extraction
of the Higgs boson properties, such as couplings and branching fractions,
which can only achieved by a comparison of theoretical predictions
with the corresponding LHC measurements.
Other examples include the determination of exclusion ranges for
specific BSM scenarios, when searches return null results,
and determination of fundamental parameters, such
as the mass of the $W$ boson.

\begin{figure}[t]
\centering
\includegraphics[width=0.49\textwidth]{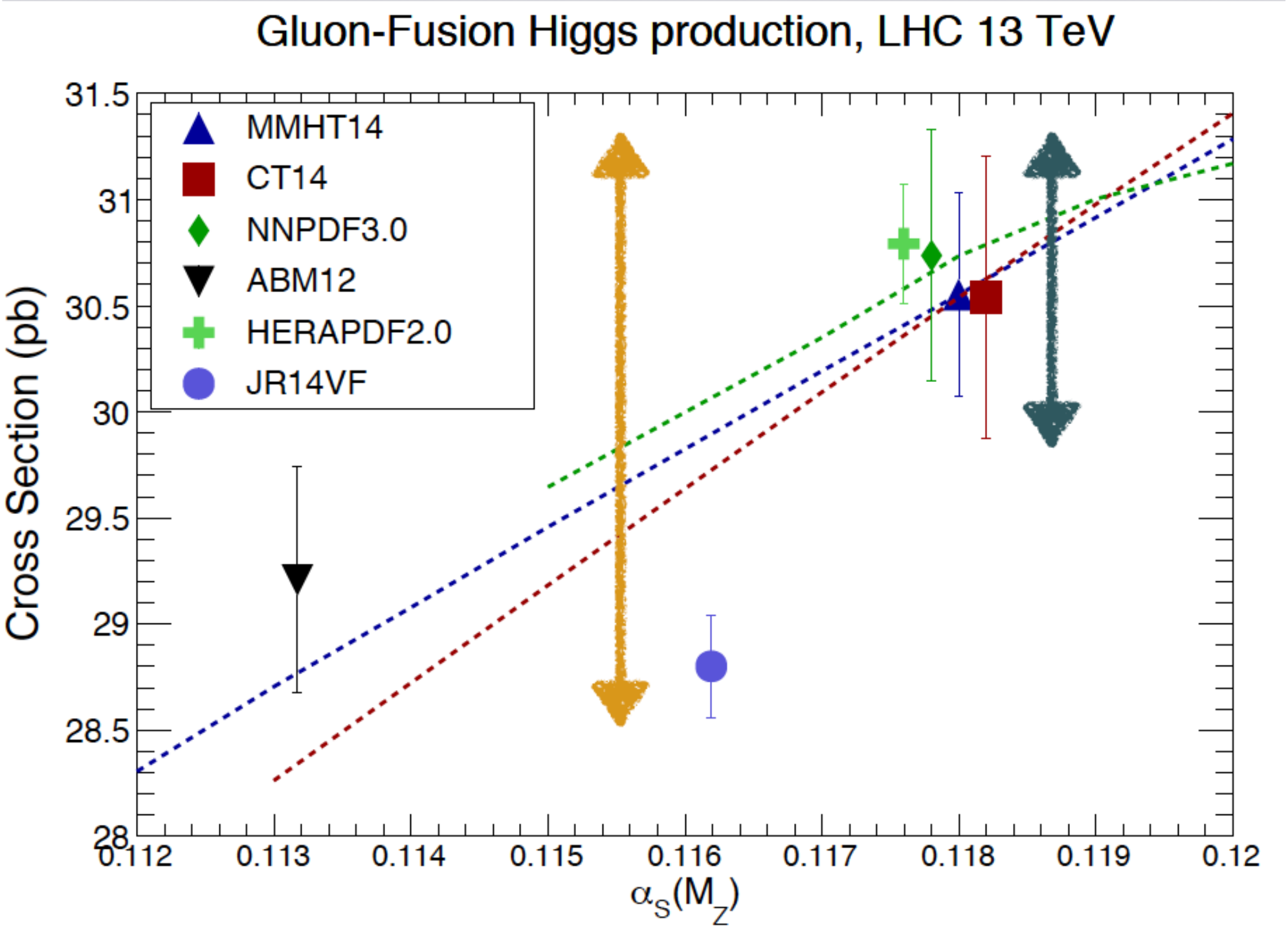}
\includegraphics[width=0.49\textwidth]{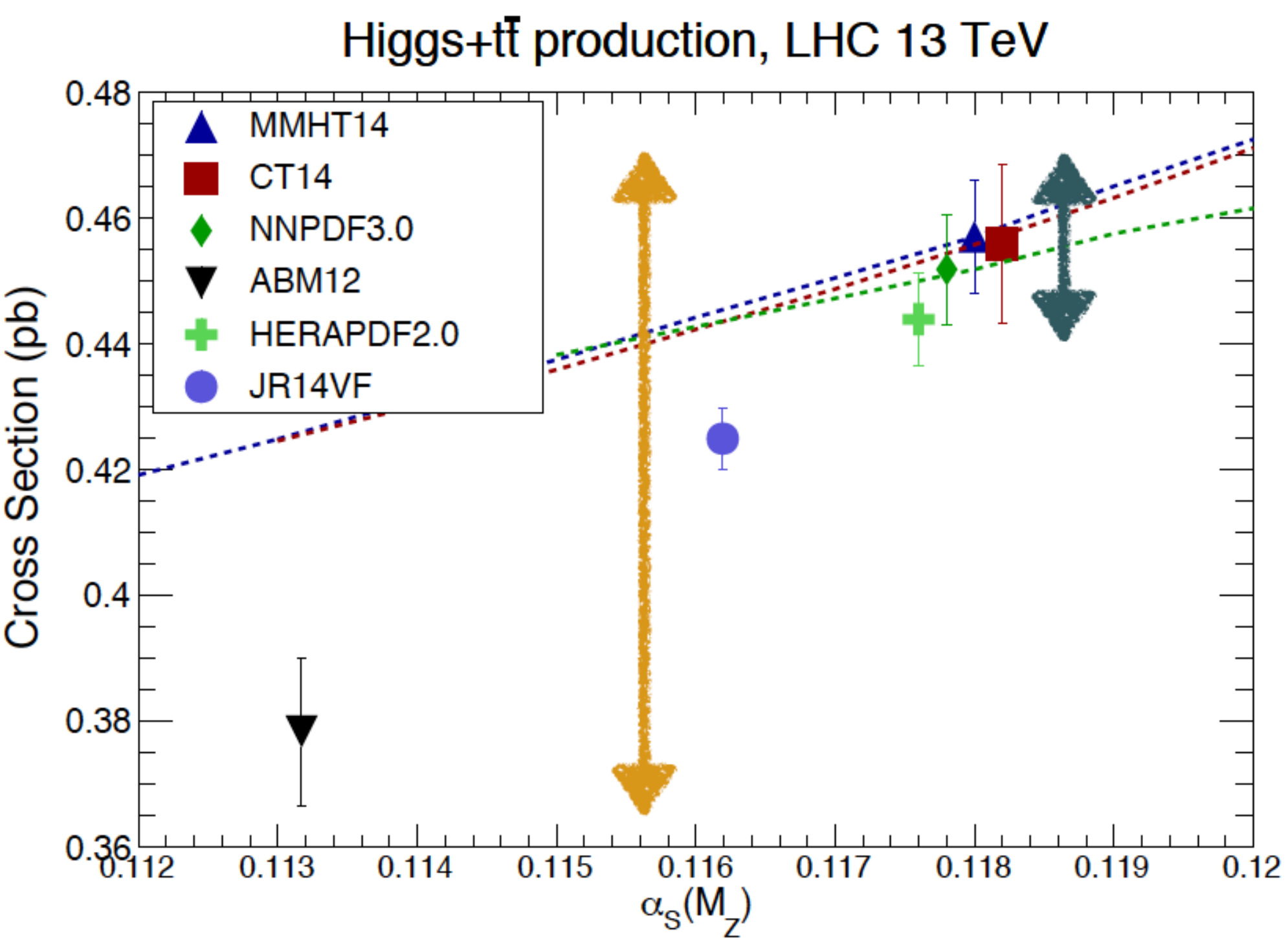}
\caption{\small NLO inclusive cross-sections, evaluated with NNLO
PDFs, for Higgs production at the LHC with $\sqrt{s}=13$ TeV
in the gluon-fusion (left)
and the $t\bar{t}$ associated production (right) channels,
for different PDFs, as a function of the native value
of the strong coupling $\alpha_s(m_Z)$ value
in each case.
In each plot, two arrows indicate schematically two possible
definitions of the {\it total PDF uncertainty} in these observables.
}
\label{higgsintro}
\end{figure}

To illustrate the challenge, in
Fig.~\ref{higgsintro} we show the NLO cross-sections (with NNLO PDFs) for Higgs production in gluon fusion and in $t\bar{t}$ associated production for different PDF sets, as a function of the native
value of the strong coupling $\alpha_s(m_Z)$.
These processes have been computed with {\tt MadGraph5\_aMC@NLO}~\cite{Alwall:2014hca,amcfast}
using default scale settings.
From Fig.~\ref{higgsintro} is clear that results from different
PDF sets are not always compatible within uncertainties.
The issue is then how one can define a total PDF+$\alpha_s$
uncertainty: this is required to extract the Higgs couplings from
the measurements of the cross-sections shown in
Fig.~\ref{higgsintro}.
Should one maybe take an envelope of the three global fits, CT14, MMHT14, and
NNPDF3.0? Or maybe one should account for
the complete spread of PDF variations?
In addition, an important motivation for having an uniform treatment
of PDF uncertainties in LHC calculations
is allowing to establish a consistent framework for the evaluation
of PDF uncertainties and their correlations in generic LHC processes.

Different points of view have been advocated to define a total PDF+$\alpha_s$
uncertainty on LHC cross-sections.
In this contribution, we review the recommendations
of the PDF4LHC Working Group~\cite{Butterworth:2015oua} for the usage of PDFs and their
uncertainties for applications at the LHC Run II.
We also briefly comment on an alternative
recommendation presented by authors
of the Ref.~\cite{Accardi:2016ndt}.

\paragraph{The PDF4LHC 2015 recommendations.}
One of the main limitations of the 2011 PDF4LHC recommendations~\cite{Botje:2011sn}
was that it required a calculation of cross-sections for each
individual PDF sets, and then combine them {\it a posteriori}
by taking the envelope of the 
PDF+$\alpha_s$ uncertainties from each set.
Moreover, the statistical interpretation of such envelope
was unclear, since it gave too much weight to outliers.
To overcome these limitations, the PDF4LHC15 recommendations
are now provided in terms of combined PDF sets.
It is important to emphasize that
switching from the
envelope of the 2011 recommendation  to the statistical combination of the
2015 one is at least in part motivated by the improved
agreement between the three PDFs sets that enter the
combination~\cite{Ball:2012wy,Butterworth:2015oua},
namely CT14~\cite{Dulat:2015mca}, MMHT14~\cite{Harland-Lang:2014zoa}
and NNPDF3.0~\cite{Ball:2014uwa}, as compared to the previous
generation sets, CT10, MSTW08 and NNPDF2.1.

\begin{figure}[t]
\centering
\includegraphics[width=0.80\textwidth]{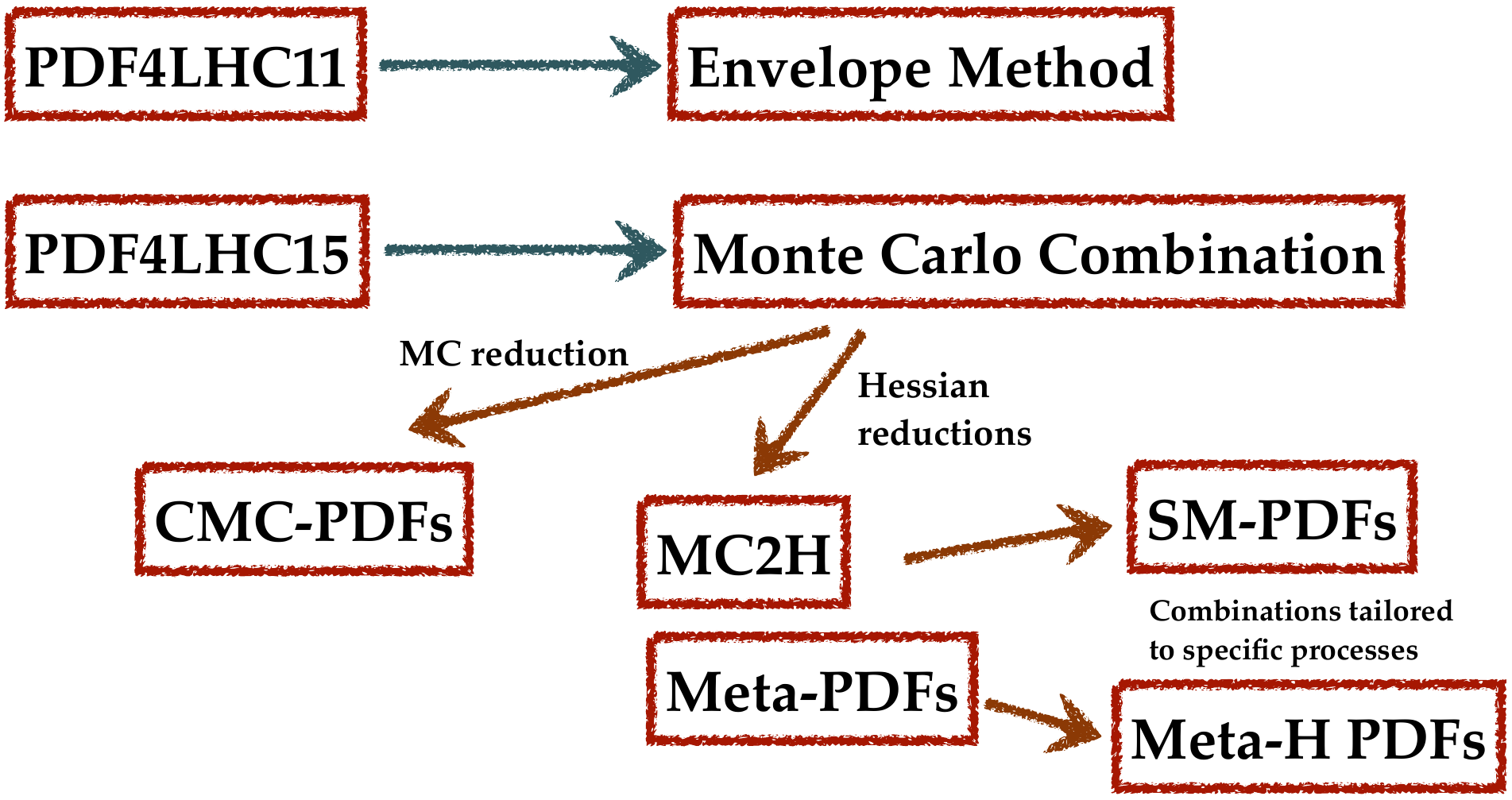}
\caption{\small Schematic representation of the different algorithms
leading to the PDF4LHC15 combined sets.
}
\label{scheme}
\end{figure}

The PDF4LHC15 combined sets,
available from
{\tt LHAPDF6}~\cite{Buckley:2014ana}, are constructed as follows.
First of all, $N_{\rm rep}=300$ Monte Carlo (MC) replicas of
NNPDF3.0
are combined with the same number from
CT14
and MMHT14 using the
Watt-Thorne method~\cite{Watt:2012tq} for the representation of Hessian
sets in terms of MC replicas.
The resulting set of $N_{\rm rep}=900$ replicas
is then reduced into more compact representations,
two Hessian ones, and a MC one.
In the latter case, the {\tt CMC-PDF} algorithm is
used~\cite{Carrazza:2015hva},
while in the former case the two Hessian reduced sets
are constructed one with $N_{\rm eig}=30$ eigenvectors,
using the {\tt META} method~\cite{Gao:2013bia},
and the other with
$N_{\rm eig}=100$ eigenvectors, using
the {\tt MC2H} algorithm~\cite{Carrazza:2015aoa}.
This procedure is summarized in Fig.~\ref{scheme}.\footnote{
Although not part of the recommendation,
it is possible to further compact the
two Hessian reduced by specifying a preferred set
of input cross-sections to be reproduced,
using either the {\tt SM-PDF}~\cite{Carrazza:2016htc} or the
{\tt META-H}~\cite{Pumplin:2009nm,Gao:2013bia}
approaches.}

\begin{figure}[t]
\centering
\includegraphics[width=0.49\textwidth]{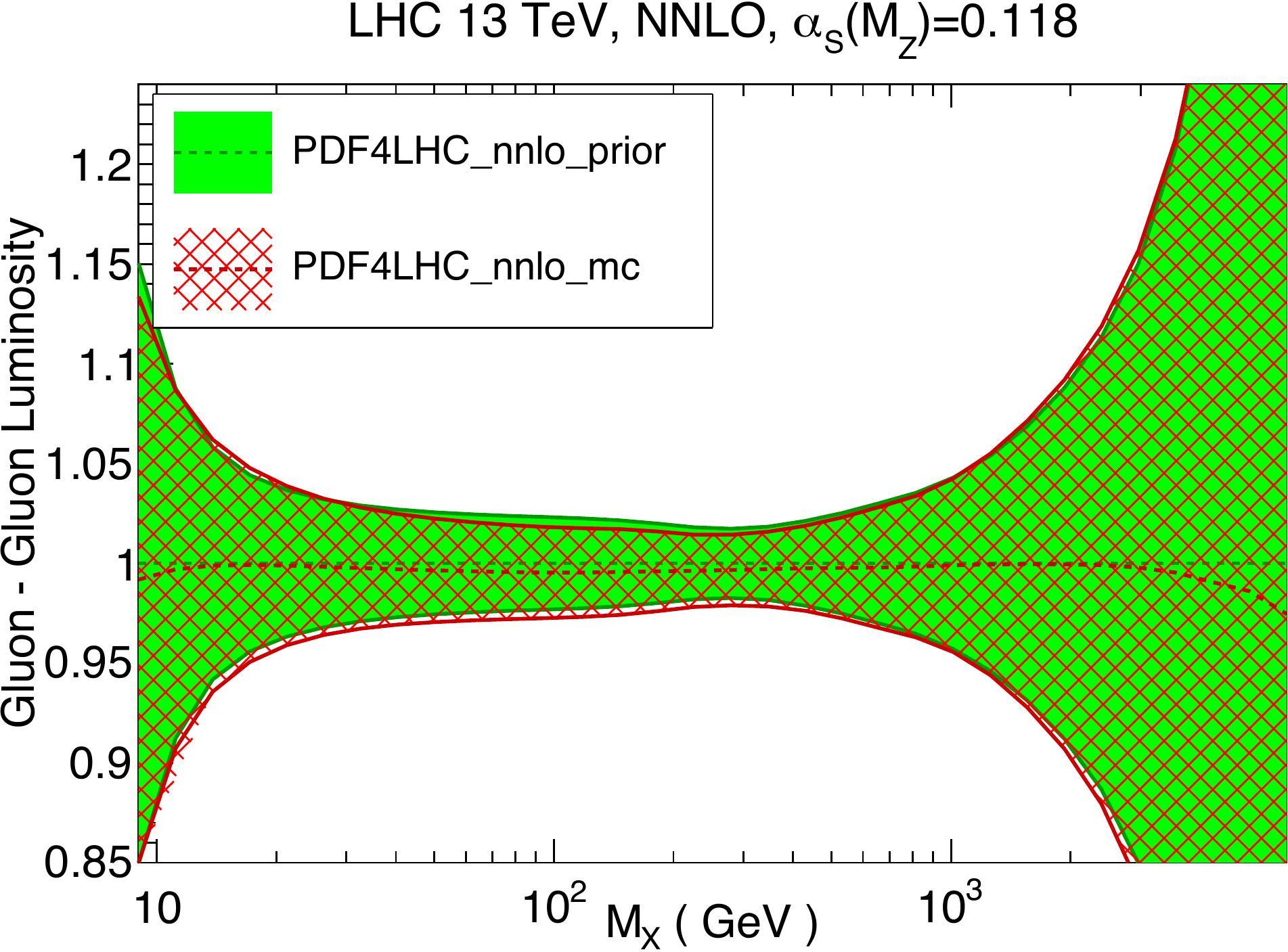}
\includegraphics[width=0.49\textwidth]{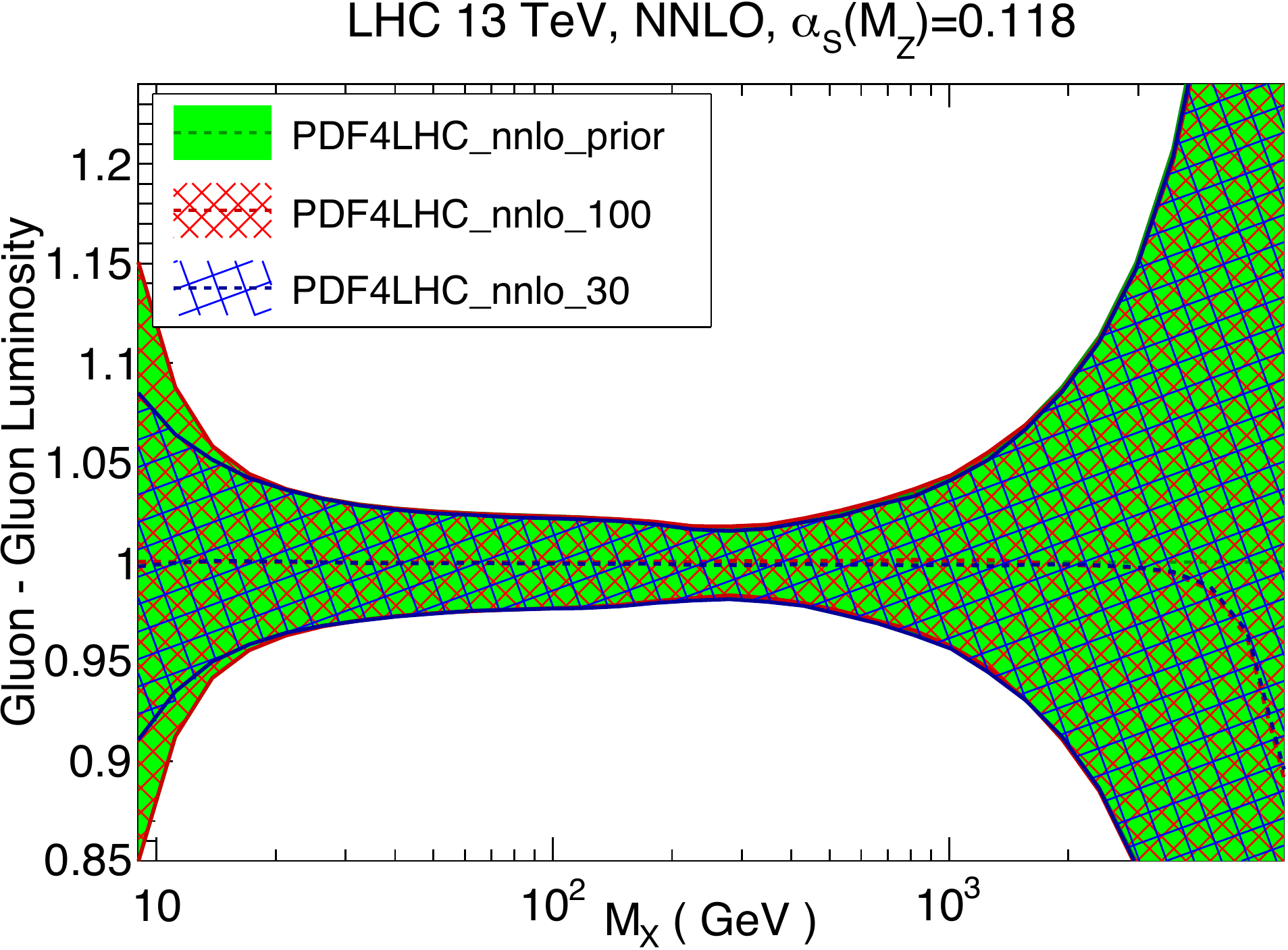}
\includegraphics[width=0.99\textwidth]{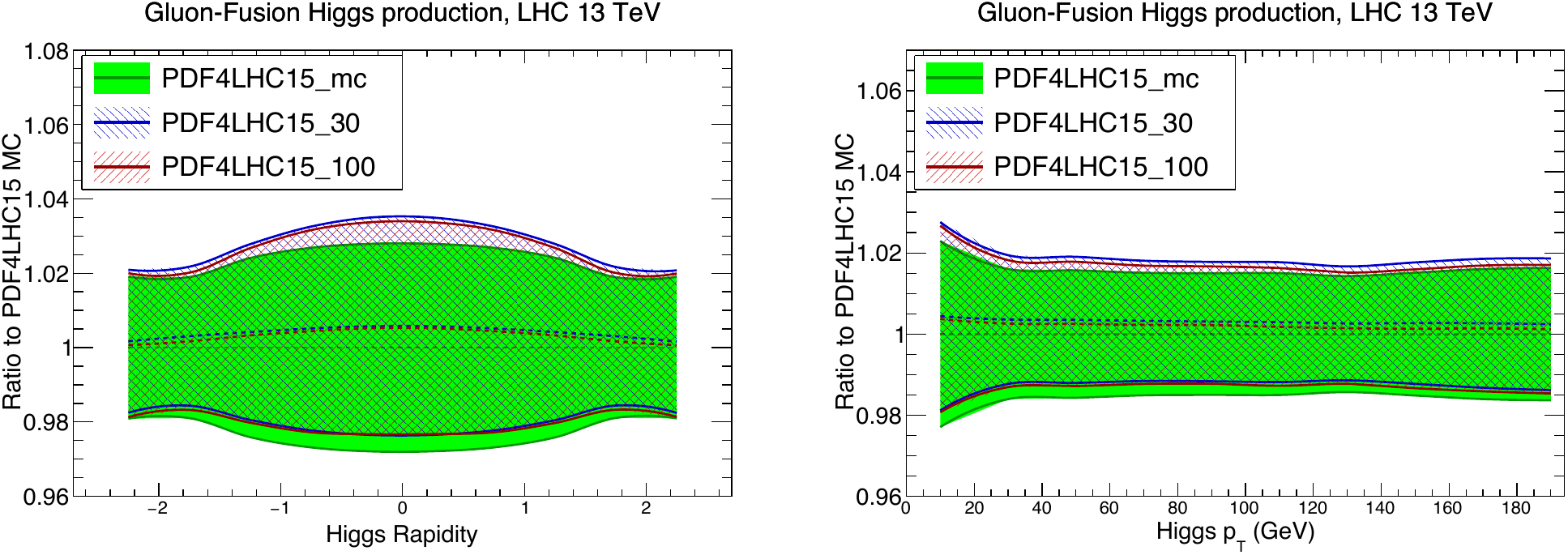}
\caption{\small Upper plots: the NNLO gluon-gluon luminosity for the
PDF4LHC15 prior combination, compared with the
subsequent Monte Carlo (left) and Hessian (right plot)
reduced sets.
Lower plots: comparison
of the predictions for differential distributions
in Higgs production in gluon fusion obtained
from the three PDF4LHC15 combined sets.
}
\label{lumi}
\end{figure}

In general, good agreement is obtained between the prior combination
and the three reduced sets in most of the relevant phase space.
In Fig.~\ref{lumi} we compare the NNLO gluon-gluon luminosity for the
PDF4LHC15 prior combination, compared with the
Monte Carlo and the Hessian 
reduced sets.
In Fig.~\ref{lumi} we also show a comparison
of the predictions for differential distributions
in Higgs production in gluon fusion at the LHC with
$\sqrt{s}=13$ TeV, in particular
the rapidity and $p_T$ distributions, obtained
from the three PDF4LHC15 combined sets.
A similar level of agreement is obtained at the level of correlations,
both between PDFs and between different collider cross-sections.

A complete set of comparisons of the PDF4LHC15 combined sets
at the level of PDFs, luminosities and
LHC cross-sections can be found at the following two websites:
\begin{center}
\url{https://www.hep.ucl.ac.uk/pdf4lhc/mc2h-gallery/website/}\\
\url{http://metapdf.hepforge.org/2016_pdf4lhc/}
\end{center}
In addition to the benchmark exercise performed
in the context of Ref.~\cite{Butterworth:2015oua}, subsequent
studies in the framework of the Les Houches
workshop~\cite{Badger:2016bpw} further explore
both the validity and the phenomenological implications of
the PDF4LHC15 recommendations, for instance addressing
in more detail the issues that arise when the prior
PDF combination exhibits non-Gaussian features.

\paragraph{An alternative recommendation.}
Recently, an alternative proposal for PDF usage at the
LHC has been advocated by Accardi {\it et al}
in Ref.~\cite{Accardi:2016ndt}.
There, a rather more conservative approach is advocated:
for precision theory predictions, the recommendation would be to use
the individual PDF sets from as many groups as possible, together
with the respective uncertainties and the values of $\alpha_s(m_Z)$,
$m_c$ and $m_b$.
In other words, the suggestion is to take the widest
possible envelope of theoretical inputs that enter
an LHC calculation.
As illustrated by Fig.~\ref{higgsintro}, adopting this recommendation
would lead to much larger theoretical uncertainties in Higgs characterization
studies and New Physics searches, affecting the physics output of
the LHC, and thus it is important to understand the reasoning
that motivates this recommendation.

In this respect, there are a number of questionable assumptions
in Ref.~\cite{Accardi:2016ndt}:
\begin{itemize}
\item It does not seem justified to disregard the wealth
of PDF-sensitive measurements available, including from the LHC~\cite{Rojo:2015acz},
for precision physics, and treat on equal footing all PDF sets
irrespectively of their level of agreement with existing
data: an envelope of results based
on fits to different-sized datasets degrades the accuracy to that of the
fit obtained from the smallest dataset.
\item In the same way as new and more
precise higher-order calculations, or Monte Carlo
simulations, replace older and less accurate ones,
also from the PDF point of view, mixing state-of-the-art PDFs with
rather older sets does not seem justified.
\item The envelope procedure leads to uncertainties that are
bigger than a statistical combination, but is justified only when
there are large and poorly understood discrepancies. This was arguably
the case at the time of the previous recommendation, but it does not
seem to be the case now.
\item We do not think it is justified to ignore the PDG average (and
the associated uncertainty)
for the strong coupling
$\alpha_s(m_Z)$~\cite{Agashe:2014kda}, implicitly declaring that
both the quoted
central value and the uncertainty are off by  a substantial
amount.
\end{itemize}

\begin{figure}[t]
\centering
\includegraphics[width=0.53\textwidth]{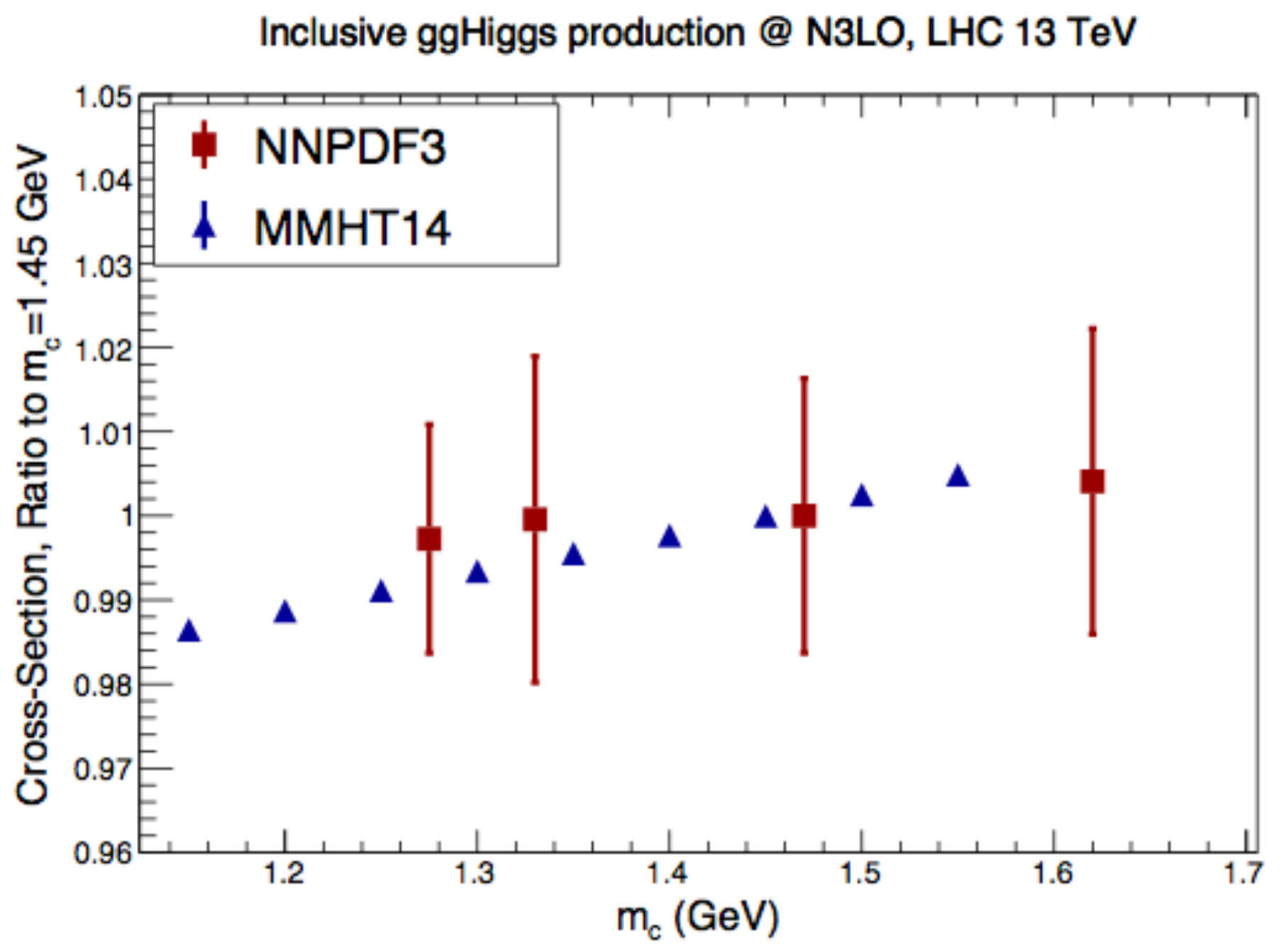}
\includegraphics[width=0.45\textwidth]{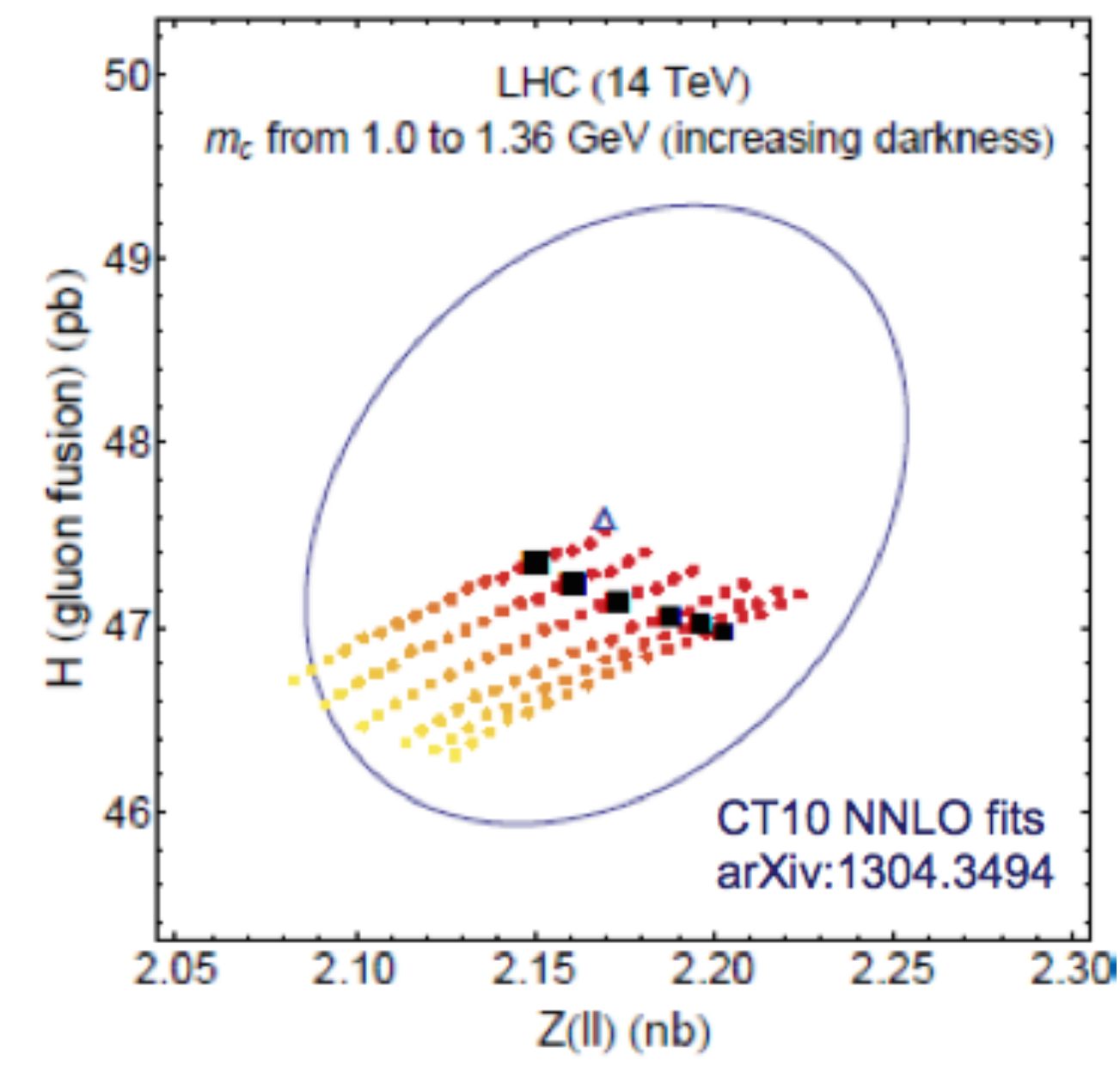}
\caption{\small Left: the Higgs cross-section in gluon
fusion computed with NLO PDFs for a range of values
of the charm pole mass $m_c$, for NNPDF3 and MMHT14.
Results are normalized to the respective values
for $m_c=1.45$ GeV.
Right: the gluon-fusion Higgs cross-section as a function of $\sigma(Z)$
for a range of values of the charm mass, from
the NNLO PDF fits of Ref.~\cite{Gao:2013wwa}.
}
\label{higgsmc}
\end{figure}

Another point raised by the authors of
Ref.~\cite{Accardi:2016ndt} is that, for the PDF fits that
enter the PDF4LHC 2015 recommendations, the numerical value of the charm
mass is effectively {\it tuned} to reach an artificial agreement
for the Higgs cross-section in gluon fusion.
A first reply to this objection is
that, in the three global
fits, the value of $\sigma(gg\to h)$
depends only mildly on the specific value of $m_c$ used.
To illustrate this point, in Fig.~\ref{higgsmc} we show
the Higgs cross-section in gluon
fusion computed with NLO PDFs for a range of values
of the charm pole mass $m_c$, for NNPDF3~\cite{Ball:2016neh} and MMHT14~\cite{Harland-Lang:2015qea}.
Even in this wide range, the cross-section varies no more
than 1\%.
In Fig.~\ref{higgsmc} we show a similar stability study,
this time in the CT10 framework~\cite{Gao:2013wwa}, using 
NNLO PDFs.
Secondly, the general-mass variable-flavour number (GM-VFN)
schemes used by the three groups have been extensively benchmarked~\cite{LHhq}
up to NNLO, and are known to differ only by small, formally subleading,
terms.
Therefore, there is little room to modify the GM-VFN matching, which
by construction is more accurate than a fixed-flavour number (FFN)
calculation.
Finally, it should be emphasized that while the conversion of the charm
mass from the pole to the $\overline{\rm MS}$ scheme
is perturbatively unstable, the same conversion is
better behaved for the bottom quark.
Together with the fact that $m^{\rm pole}_b-m^{\rm pole}_c$
is free of renormalon ambiguities, this
implies that a measurement
of $m_b^{\overline{\rm MS}}$ leads to a reasonably accurate
prediction of $m^{\rm pole}_c=1.51 \pm 0.13$~\cite{Bauer:2004ve},
consistent
with the values used in the global fits.

\paragraph{Outlook.}
The PDF4LHC 2015 recommendations are the result of a joint
effort by theorists and experimentalists
aiming to provide a
robust estimate of the combined PDF+$\alpha_s$
uncertainties for precision LHC calculations.
The main forum of the PDF4LHC Working Group are
its periodic meetings,
which provide a unique opportunity for the cross-talk between
theory and experiment, as well as between PDF fitters.
Topics that will be explored in the coming months include
the impact of LHC data from the 13 TeV runs,
including new NNLO calculations
in PDF fits and the role of electroweak corrections
and the photon PDF.
Eventually, PDF4LHC will present updated recommendations to take into
account developments from the theory, data,
and methodology points of view.
For instance, future updates
might include additional PDF sets.
In any case, the general combination strategy developed in
the context of present recommendation is flexible and robust enough
to accommodate these and other foreseeable updates.

\paragraph{Acknowledgments.}
I am grateful to my colleagues of the PDF4LHC Working Group
and in particular to the authors of~\cite{Butterworth:2015oua}
for innumerable illuminating discussions.
This work has been supported by an STFC Rutherford Fellowship
and Grant ST/K005227/1 and ST/M003787/1, and
by an European Research Council Starting Grant ``PDF4BSM''.


\begin{thebibliography}{10}

\bibitem{Forte:2013wc}
S.~Forte and G.~Watt, {\it {Progress in the Determination of the Partonic
  Structure of the Proton}},  {\em Ann.Rev.Nucl.Part.Sci.} {\bf 63} (2013) 291,
  [\href{http://arxiv.org/abs/1301.6754}{{\tt arXiv:1301.6754}}].

\bibitem{Alwall:2014hca}
J.~Alwall, R.~Frederix, S.~Frixione, V.~Hirschi, F.~Maltoni, et~al., {\it {The
  automated computation of tree-level and next-to-leading order differential
  cross sections, and their matching to parton shower simulations}},  {\em
  JHEP} {\bf 1407} (2014) 079, [\href{http://arxiv.org/abs/1405.0301}{{\tt
  arXiv:1405.0301}}].

\bibitem{amcfast}
V.~Bertone, R.~Frederix, S.~Frixione, J.~Rojo, and M.~Sutton, {\it {aMCfast:
  automation of fast NLO computations for PDF fits}},  {\em JHEP} {\bf 1408}
  (2014) 166, [\href{http://arxiv.org/abs/1406.7693}{{\tt arXiv:1406.7693}}].

\bibitem{Butterworth:2015oua}
J.~Butterworth et~al., {\it {PDF4LHC recommendations for LHC Run II}},  {\em J.
  Phys.} {\bf G43} (2016) 023001, [\href{http://arxiv.org/abs/1510.03865}{{\tt
  arXiv:1510.03865}}].

\bibitem{Accardi:2016ndt}
A.~Accardi et~al., {\it {Recommendations for PDF usage in LHC predictions}},
  \href{http://arxiv.org/abs/1603.08906}{{\tt arXiv:1603.08906}}.

\bibitem{Botje:2011sn}
M.~Botje et~al., {\it {The PDF4LHC Working Group Interim Recommendations}},
  \href{http://arxiv.org/abs/1101.0538}{{\tt arXiv:1101.0538}}.

\bibitem{Ball:2012wy}
R.~D. Ball, S.~Carrazza, L.~Del~Debbio, S.~Forte, J.~Gao, et~al., {\it {Parton
  Distribution Benchmarking with LHC Data}},  {\em JHEP} {\bf 1304} (2013) 125,
  [\href{http://arxiv.org/abs/1211.5142}{{\tt arXiv:1211.5142}}].

\bibitem{Dulat:2015mca}
S.~Dulat, T.-J. Hou, J.~Gao, M.~Guzzi, J.~Huston, P.~Nadolsky, J.~Pumplin,
  C.~Schmidt, D.~Stump, and C.~P. Yuan, {\it {New parton distribution functions
  from a global analysis of quantum chromodynamics}},  {\em Phys. Rev.} {\bf
  D93} (2016), no.~3 033006, [\href{http://arxiv.org/abs/1506.07443}{{\tt
  arXiv:1506.07443}}].

\bibitem{Harland-Lang:2014zoa}
L.~A. Harland-Lang, A.~D. Martin, P.~Motylinski, and R.~S. Thorne, {\it {Parton
  distributions in the LHC era: MMHT 2014 PDFs}},  {\em Eur. Phys. J.} {\bf
  C75} (2015) 204, [\href{http://arxiv.org/abs/1412.3989}{{\tt
  arXiv:1412.3989}}].

\bibitem{Ball:2014uwa}
{\bf NNPDF} Collaboration, R.~D. Ball et~al., {\it {Parton distributions for
  the LHC Run II}},  {\em JHEP} {\bf 04} (2015) 040,
  [\href{http://arxiv.org/abs/1410.8849}{{\tt arXiv:1410.8849}}].

\bibitem{Buckley:2014ana}
A.~Buckley, J.~Ferrando, S.~Lloyd, K.~Nordström, B.~Page, et~al., {\it
  {LHAPDF6: parton density access in the LHC precision era}},  {\em
  Eur.Phys.J.} {\bf C75} (2015) 132,
  [\href{http://arxiv.org/abs/1412.7420}{{\tt arXiv:1412.7420}}].

\bibitem{Watt:2012tq}
G.~Watt and R.~S. Thorne, {\it {Study of Monte Carlo approach to experimental
  uncertainty propagation with MSTW 2008 PDFs}},  {\em JHEP} {\bf 1208} (2012)
  052, [\href{http://arxiv.org/abs/1205.4024}{{\tt arXiv:1205.4024}}].

\bibitem{Carrazza:2015hva}
S.~Carrazza, J.~I. Latorre, J.~Rojo, and G.~Watt, {\it {A compression algorithm
  for the combination of PDF sets}},  {\em Eur. Phys. J.} {\bf C75} (2015) 474,
  [\href{http://arxiv.org/abs/1504.06469}{{\tt arXiv:1504.06469}}].

\bibitem{Gao:2013bia}
J.~Gao and P.~Nadolsky, {\it {A meta-analysis of parton distribution
  functions}},  {\em JHEP} {\bf 1407} (2014) 035,
  [\href{http://arxiv.org/abs/1401.0013}{{\tt arXiv:1401.0013}}].

\bibitem{Carrazza:2015aoa}
S.~Carrazza, S.~Forte, Z.~Kassabov, J.~I. Latorre, and J.~Rojo, {\it {An
  Unbiased Hessian Representation for Monte Carlo PDFs}},  {\em Eur. Phys. J.}
  {\bf C75} (2015), no.~8 369, [\href{http://arxiv.org/abs/1505.06736}{{\tt
  arXiv:1505.06736}}].

\bibitem{Carrazza:2016htc}
S.~Carrazza, S.~Forte, Z.~Kassabov, and J.~Rojo, {\it {Specialized minimal PDFs
  for optimized LHC calculations}},  {\em Eur. Phys. J.} {\bf C76} (2016),
  no.~4 205, [\href{http://arxiv.org/abs/1602.00005}{{\tt arXiv:1602.00005}}].

\bibitem{Pumplin:2009nm}
J.~Pumplin, {\it {Data set diagonalization in a global fit}},  {\em Phys. Rev.}
  {\bf D80} (2009) 034002, [\href{http://arxiv.org/abs/0904.2425}{{\tt
  arXiv:0904.2425}}].

\bibitem{Badger:2016bpw}
J.~R. Andersen et~al., {\it {Les Houches 2015: Physics at TeV Colliders
  Standard Model Working Group Report}},  in {\em {9th Les Houches Workshop on
  Physics at TeV Colliders (PhysTeV 2015) Les Houches, France, June 1-19,
  2015}}, 2016.
\newblock \href{http://arxiv.org/abs/1605.04692}{{\tt arXiv:1605.04692}}.

\bibitem{Rojo:2015acz}
J.~Rojo et~al., {\it {The PDF4LHC report on PDFs and LHC data: Results from Run
  I and preparation for Run II}},  {\em J. Phys.} {\bf G42} (2015) 103103,
  [\href{http://arxiv.org/abs/1507.00556}{{\tt arXiv:1507.00556}}].

\bibitem{Agashe:2014kda}
{\bf Particle Data Group} Collaboration, K.~Olive et~al., {\it {Review of
  Particle Physics}},  {\em Chin.Phys.} {\bf C38} (2014) 090001.

\bibitem{Ball:2016neh}
{\bf The NNPDF} Collaboration, R.~D. Ball, V.~Bertone, M.~Bonvini, S.~Carrazza,
  S.~Forte, A.~Guffanti, N.~P. Hartland, J.~Rojo, and L.~Rottoli, {\it {A
  Determination of the Charm Content of the Proton}},
  \href{http://arxiv.org/abs/1605.06515}{{\tt arXiv:1605.06515}}.

\bibitem{Harland-Lang:2015qea}
L.~A. Harland-Lang, A.~D. Martin, P.~Motylinski, and R.~S. Thorne, {\it {Charm
  and beauty quark masses in the MMHT2014 global PDF analysis}},  {\em Eur.
  Phys. J.} {\bf C76} (2016), no.~1 10,
  [\href{http://arxiv.org/abs/1510.02332}{{\tt arXiv:1510.02332}}].

\bibitem{Gao:2013wwa}
J.~Gao, M.~Guzzi, and P.~M. Nadolsky, {\it {Charm quark mass dependence in a
  global QCD analysis}},  {\em Eur.Phys.J.} {\bf C73} (2013) 2541,
  [\href{http://arxiv.org/abs/1304.3494}{{\tt arXiv:1304.3494}}].

\bibitem{LHhq}
J.~Rojo et~al., ``{Chapter 22 in: J.~R.~Andersen et al., "The SM and NLO
  multileg working group: Summary report"}.'' arXiv:1003.1241, 2010.

\bibitem{Bauer:2004ve}
C.~W. Bauer, Z.~Ligeti, M.~Luke, A.~V. Manohar, and M.~Trott, {\it {Global
  analysis of inclusive B decays}},  {\em Phys. Rev.} {\bf D70} (2004) 094017,
  [\href{http://arxiv.org/abs/hep-ph/0408002}{{\tt hep-ph/0408002}}].

\end{thebibliography}
\providecommand{\href}[2]{#2}\begingroup\raggedright\endgroup

\end{document}